# Contra generative AI detection in higher education assessments

Cesare Giulio Ardito, Lecturer, Department of Mathematics, School of Natural Sciences, Faculty of Science and Engineering, University of Manchester, Manchester, United Kingdom. Email: cesaregiulio.ardito@manchester.ac.uk .

**Abstract**

This paper presents a critical analysis of generative Artificial Intelligence (AI) detection tools in higher education assessments. The rapid advancement and widespread adoption of generative AI, particularly in education, necessitates a reevaluation of traditional academic integrity mechanisms. We explore the effectiveness, vulnerabilities, and ethical implications of AI detection tools in the context of preserving academic integrity. Our study synthesises insights from various case studies, newspaper articles, and student testimonies to scrutinise the practical and philosophical challenges associated with AI detection. We argue that the reliance on detection mechanisms is misaligned with the educational landscape, where AI plays an increasingly widespread role. This paper advocates for a strategic shift towards robust assessment methods and educational policies that embrace generative AI usage while ensuring academic integrity and authenticity in assessments.

**Keywords:** Generative AI, AI Detection, Assessment, Higher Education, Academic Integrity, Higher Education Policy.

## 1. Introduction

The most basic form of detection of content generated by AI consists of simple human insight, when someone believes that the text they are reading has been generated, completely or in part, by AI. In some cases, the evidence is overwhelming: for instance, the author once received a reference letter ending with the words "regenerate response", which appeared at the end of every ChatGPT response generated before September 2023.

However, in the absence of unambiguous signs, studies show that humans are overconfident in their ability to detect AI-generated content [7], [28], [26] both in general and in specialised contexts [18]. In higher education, therefore, it is not wise for an educator to rely on their own ability to differentiate AI-generated text from human-written text.

The rapid mass availability and enhanced capabilities of generative AI tools, starting with ChatGPT's release to the public in November 2022, gave rise to a multitude of machine-aided detectors: text-classifier tools that output the probability that a given textual input was generated by artificial intelligence. The underlying idea is simple and appealing: text is pasted in a box, and the detector can recognise whether it has been generated by AI, through several statistical evaluations such as checking the sequence of words (tokens) against the weights of the model. Such detectors require no technical expertise and are easily accessible to the public.

In early 2023, many competing solutions were put on the market, either by established companies like Turnitin or by focused new startups like CopyLeaks, GptZero, Sapling, ZeroGPT and many others. A further text classifier tool was released by OpenAI, and later withdrawn due to its low rate of accuracy, with the company choosing to only focus on detection of AI-generated audio and visual content in the future [37]. In general, AI detection in academic settings has emerged as a highly controversial topic, which has received considerable attention.

In this paper, we critically examine the viability and implications of employing AI detection tools in the realm of academic integrity. Given the rapid advancement and widespread availability of generative AI, particularly in the context of higher education, there is an urgent need to evaluate the effectiveness and appropriateness of these detection methods. Our analysis delves into the broader considerations surrounding AI detection, questioning its place in future academic settings. We examine the accuracy, vulnerabilities, and ethical dimensions of AI detection tools, assessing the implications of incorporating such technology in an academic setting

First, we undertake a comprehensive examination of the role and efficacy of AI detection tools within the context of academic integrity, anchoring our analysis in a variety of case studies, newspaper articles, and student testimonies. We then examine AI detection from a broader perspective, highlighting fundamental issues beyond the current shortcomings of the technology, and reflecting on whether it has a place in the future in academic contexts where AI usage is integrated and part of the learning objectives. Finally, we argue in favour of the development of an alternative framework, intending to teach critical and ethical usage of generative AI to students, while safeguarding academic integrity and the authenticity of assessments.

## 2. The pitfalls in detecting generative AI output

AI detection tools are based on a variety of statistical approaches and show varying performances. It is not within the scope of this paper to discuss the technical details of each – rather, we focus on the opportunity and appropriateness of employing detection solutions. To the interested reader, we recommend the following survey paper [54]. However, it is still beneficial to our discussion to highlight common issues that have emerged, together with potential pitfalls, as it is undeniable that the case for using AI detectors would be stronger if they were more reliable.

*Vulnerabilities*

AI detectors can be vulnerable to several attacks. Perhaps unsurprisingly, paraphrasing attacks (consisting in the replacement of some words in the text with synonyms) have been shown to successfully bypass several AI detectors [51][29], and paraphrasing tools such as Quillbot or GPT-Minus1 are widespread.

Moreover, simple prompting strategies such as asking for a certain style, or guiding the model towards one, have been shown to lower the sensitivity of detectors [14], [15]. An example prompt was the addition of "*Include personal reflections, use a mix of long and short sentences, employ rhetorical questions to engage the reader, maintain a conversational tone in parts, and play around with the paragraph structure to create a dynamic and engaging piece of writing*" at the end of the prompt. Another strategy was asking the model to regenerate a new answer, adding further requirements.

Parameter tweaking, such as changing the temperature or the top P score of the model, adjusts how the model generates text, reducing the sensitivity of detectors [15]. It is worth highlighting that this is likely due to the underrepresentation of text generated under altered parameters in the detector's AI-generated training set, which shows that detectors focus on particular incarnations of the model, and their performance can degrade when any significant deviation from its default setup is introduced.

Further, several commercial services allow users to input their text and check its score against multiple AI detectors, enabling a trial-and-error process that, through paraphrasing, rewriting or other means, ends up producing seemingly human-written text.

A more general argument is that any vulnerability of AI detection mechanisms automatically gives rise to a critical issue, undermining their reliability and effectiveness. Indeed, the core of AI technology is automation: as soon as a vulnerability that reduces the likelihood of the output being identified as AI-generated is found, a generative model can be prompted to reproduce the vulnerability and be instantly released to the public. For instance, it would be easy to program a version of GPT that follows styling guidelines similar to the ones above for each user-written prompt.

In the context of academic integrity, the existence of online networks dedicated to facilitating academic malpractice is well-documented [36], [39], so a vulnerability could be disseminated and widely exploited in a matter of days. The combination of this rapid dissemination with the commercial incentives to identify such weaknesses makes reliance on AI detection tools for maintaining academic integrity a fundamentally flawed strategy.

*Future-proof detection*

It's a straightforward exercise to demonstrate that any sentence can be part of the output of a Large Language Model (LLM).

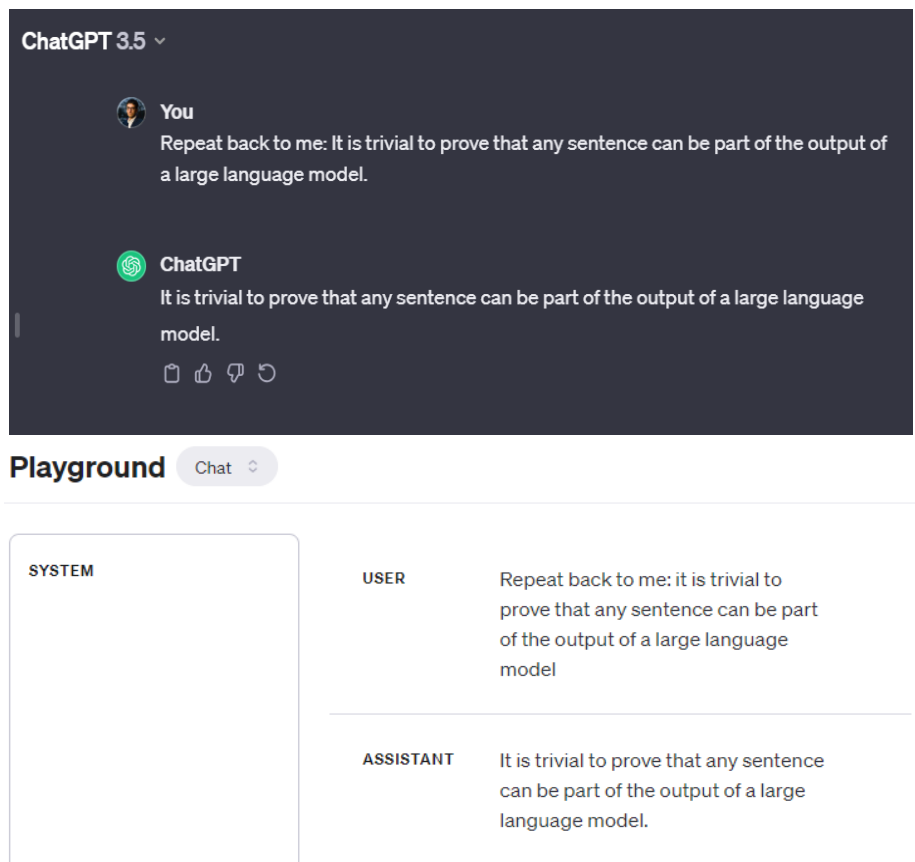

Conversation with ChatGPT [3] and with gpt-3.5-turbo (default parameters) on the OpenAI playground

This is less trivial than it may appear at first glance, as LLMs generate tokens one at a time, with no further insight or planning [8] – in other words, the model is *writing* those words, not copying and pasting the whole string back. Oversimplifying, the current limited effectiveness of detectors is due to the fact that LLMs, given their training and architecture, are more likely to write certain sentences than others [54]. Indeed, LLMs merely use their training dataset to produce statistically likely sentences. However, as LLMs continue to advance in complexity and capability, their ability to make use of more complex syntactic structures and the information in their training datasets increases, so the scope of their text generation grows. This implies that, over time, these models might attain the proficiency to generate any form of text, in any style, effectively and successfully mimicking the entirety of human writing output.

Asymptotically, the datasets used to test the efficacy of AI detection tools could potentially expand to encompass all conceivable text, making the very concept of detection paradoxical, similar to a search in Jorge Luis Borges' Library of Babel [4]. We are still far from such a scenario, but the speed of improvements is itself concerning: for instance, GPT-4 has a richer prose than GPT-3.5 [14], and many detectors perform significantly worse on the former. These trends are likely to continue, so distinguishing between AI and human authorship becomes a perpetual cat-and-mouse game, with detection always a step behind, assuming it could even catch up.

Consequently, relying on AI detection to maintain academic integrity is doomed to become increasingly untenable as time passes. Instead, it is imperative to design educational systems and assessment methods that do not depend on the flawed premise of being able to accurately distinguish AI-generated content from human work. This shift is not just a practical necessity but a strategic adaptation to the rapidly evolving landscape of AI capabilities, ensuring that educational integrity is upheld through more robust and future-proof methods

*Bias and non-English speakers*

Even if very high accuracy scores were possible to achieve against current large language models, it may still occur for false positives to not be uniformly distributed across text samples, but instead to disproportionately affect certain subgroups of human writers. Recently, it has been reported that AI detectors were more likely to incorrectly flag papers from non-native English speakers as AI-generated, based on a study from Stanford [31]. To the author's knowledge, at the moment the only other study on the subject appeared on Turnitin's own blog [1], beginning with a critique of [31] as largely based on a small sample of 91 works below 150 words. Further, Turnitin's article reported that their AI detector showed no such bias when tested on larger databases of longer text, although it must be noted that Stanford's original study did not include this detector. We conclude that, at the moment, there is not enough evidence to assert the presence of such bias in AI detectors.

However, showing bias against non-native English speakers would not be surprising from an intuitive point of view: several detectors examined in the paper employ a measure of *perplexity*. A large language model can be thought of as a stochastic Markov chain, whose probabilities (weights) come from training on a big corpus of human-written text, so a model is more likely to write "common" sentences, i.e. those that follow patterns with widespread usage. Again oversimplifying, perplexity measures how surprising it would be if the model produced a given sentence. English speakers that aren't native speakers are likely to adhere more closely to standard syntax in their prose, and hence exhibit lower perplexity scores and trigger a higher rate of false positives with AI output detectors. An extreme example of this dynamic appeared in many news reports in July 2023: the constitution of the United States of America is flagged as AI-generated by several detectors [11] since its perplexity score is extremely low as it appears throughout every model's training data, and is reproduced identically in most instances when it is cited.

Measuring perplexity is not the only approach, and some detectors use alternative measurements, often combined with machine learning [10]. However, the example of perplexity measurement is a cautionary tale: an overall low false positivity rate is, on its own, not enough to promote the widespread usage of an AI detector. Indeed, the statistical nature of detection methods may cause disproportionate false positivity rates to accumulate against certain population subgroups, based on characteristics of their prose. Strong, robust statistical analyses of such tools should be performed before these tools can be accepted as reliable, but given the rapid advancements in generative AI, it is unlikely that any tool could be scrutinised this closely before it becomes obsolete.

*Dataset pollution*

Accuracy measures for detectors are taken by comparing a dataset of AI-generated essays with another of human-generated essays. The latter is normally made of essays, papers or articles that were written before the release of ChatGPT or even GPT-2, to avoid AI-written or co-created content [43]. Despite Generative AI being a very recent phenomenon, the unreliability of any datasets that contain post-2023 data as authentic human data is widely accepted [23].

This has implications for the future reliability of generative AI models, which will train on content that will increasingly include AI-generated pieces. It also affects detection: on one hand, it will become harder to source authentic human-written datasets, and retraining detectors on content that they self-authenticate as human-written could amplify existing bias and inaccuracies. On the other hand, language evolves with time, so detectors trained only on old pre-2023 text may start misclassifying human-written text as AI-generated simply because it is anachronistic. Once more, this highlights that AI detection can at most be a stop-gap short-term solution, and that more fundamental changes are necessary to safeguard future academic integrity.

*Watermarking*

An interesting, alternative approach to enable detection is watermarking: this consists of artificially embedding a hidden "signature" in text generated by a model, that can later be detected. A toy example would be as follows: dictionary words are divided into two sets (red and blue), and the model chooses words alternating between the two. For sufficiently long paragraphs, it would be extremely unlikely for a human author to have picked words respecting the alternating sequence, enabling successful detection. While this would still be vulnerable to paraphrasing attacks, more sophisticated approaches have been proposed that are more resistant to standard attacks. Nevertheless, a recent paper claims to have proved that no watermark is immune to obfuscation [55] through paraphrasing.

There is a further issue: current models are not watermarked, and they are already available in a multitude of solutions. Further, not all of them can be withdrawn, since some models have been leaked online [49]. Should watermarking become a mainstream solution for AI detection, perhaps even enforced by regulation, there would be a clear business incentive in having non-watermarked models available for a variety of applications (among which contract cheating is certainly included). In other words, it appears that the cat is out of the bag, and watermarking, feasible or not, is not an advisable path forward.

# 3. Detectors are not useful

We have seen how current detectors possess several vulnerabilities, and how future detectors are likely to also have significant vulnerabilities and shortcomings. In this section, we'll argue that any approach that incorporates AI detection comes with problems beyond technological shortcomings.

*Detectors are misaligned*

We start by pointing out how AI detection and traditional technological aids differ crucially from academic malpractice. Plagiarism detectors such as Turnitin Similarity Report [44] have proven effective in identifying non-original text and tracing its sources, facilitating the detection of plagiarism in academic assignments. Users of such tools want to check that the text has not been copied from another source, and the tool is perfectly aligned with this goal; further, the user is provided with a report and the means to independently verify its accuracy, through links to the alleged plagiarised sources.

AI detectors operate on a fundamentally different principle. They assess the degree to which text aligns with the most common outputs of supported large language models and do not provide any means to independently verify their assessment. Moreover, the primary concern for users is to discern the student's original contribution, which does not necessarily correlate with the portion of text that has likely been generated by AI. This discrepancy becomes especially problematic when considering legitimate uses of AI in academic work, such as reviewing, paraphrasing, enhancing expression, or through accessibility tools.

For instance, consider two hypothetical students: Alice, who writes an original draft essay and then uses an LLM for refinement, implementing many of the suggestions; and Bob, who generates an essay with ChatGPT a few minutes before his deadline, changing a few words, restructuring a few sentences and then submitting. Alice's work would be more likely to receive a high score on a detector since the last form of the work comes directly from a large language model – on the other hand, Bob's work is less likely to be similar to AI output. Hence, the detector would not accurately reflect the extent of their individual contributions or the presence of malpractice.

Further, for the majority of courses, Alice would still have satisfied the learning objectives that she was meant to achieve. Moreover, it can be argued that even if writing skills were being assessed, she would still have learned how to express her viewpoints in a professional, coherent and formal way, and she would be able to replicate said performance in any future instance, for example in the employment sector. Indeed, generative AI is here to stay – we do not live in a world where large language models cannot help us in a variety of tasks. That world is gone, and it will not come back.

*Detectors are not falsifiable*

There is an inherent challenge in validating the results of AI detectors. If a detector indicates that a substantial portion of a text is AI-generated, this claim is not verifiable, as the detector provides no evidence other than the results of its assessment. Nor can such proof exist – as we have seen, it is possible to make an LLM output any given text, so even the explicit provision of the used prompts through reverse prompt engineering could not possibly prove anything.

The wider availability and the aggressive marketing of detection tools, targeting academics and teachers, has in several instances impacted students, who were accused of malpractice they did not commit, and had no way to acquit themselves. Moreover, the integration of AI detection into mainstream anti-plagiarism tools like Turnitin has further complicated the landscape, drawing a false parallel between two very different tools[1]. The resultant academic environment, where students and academics were left to interpret "AI scores" without a coherent policy framework, led to inconsistent and unjust outcomes. In April 2023, a student was accused of malpractice due to misdetection and was only able to prove her innocence thanks to Google Docs' autosaving drafts functionality [27]. Another student, accused similarly, had to access expensive legal advice and consultancy from a linguistics expert to have the case dropped [42]. A simple search on social media – Instagram, Reddit, TikTok – reveals thousands of reports of similar cases across all ages and levels [20]. In August, the Washington Post published a guide for students to defend themselves from wrong accusations due to misdetection [16].

Among advice for students to defend themselves, two strategies commonly appear: the first is to keep drafts of their work, ideally on an online repository that can certify dates, so they can show the evolution of the work in case they are accused of malpractice. The second is to independently check their scores on AI detectors before submission, through the use of several commercial services that claim to replicate the performance of mainstream AI detectors [35]. The latter option is, of course, terrible for privacy and integrity, as these tools and their privacy policies are often opaque and unreliable: it is not inconceivable to hypothesise a scenario when a human-written essay is uploaded through one of these portals from the student, and then resold by an essay mill to a third party, leading to a hard-to-dismiss plagiarism accusation against the student validated through Turnitin's plagiarism detector!

In general, being subject to an academic malpractice investigation is known to induce severe anxiety in students, due to the formality of the processes and the severe consequences that can result from the verdict [21], [24]. A wrongful malpractice accusation can be likened to the famous opening of Kafka's "The Trial," where Josef K. faces the consequences of inexplicable, unfalsifiable accusations, and has no way to prove himself innocent while thrust into a daunting and formalised process of scrutiny. This can be deeply unsettling and contribute to a sense of helplessness and anxiety.

This situation is further compounded by the existing challenges to student mental health. The contemporary academic environment, characterised by high demands, competitive pressures, and the inherent challenges of balancing academic and personal life, already places a significant mental burden on students. Introducing the added stressor of potential wrongful accusations of AI-assisted malpractice, especially with the increasing deployment of imperfect AI detection tools, risks exacerbating these mental health challenges. Moreover, repeated occurrences of false malpractice accusations could erode trust between students and educational institutions, a cornerstone of the educational process, which could have far-reaching implications.

Even a detector with a seemingly low false positive rate of 1% poses a dramatic threat of wrongful accusations when considering the volume of essays a student writes throughout their academic career. The number of written submissions, including coursework, projects, reports, and dissertations, varies by discipline, often ranging up to 70 pieces or more. A quick back-of-the-envelope calculation reveals that, assuming a uniform false positive rate of only 1% across all submissions, an entirely honest student who completes 70 essays during their degrees faces a 50.5% chance of being erroneously flagged for AI use at least once. Extrapolating this to the scale of a typical educational institution, with its vast student body, such a detection system could lead to thousands of unfounded allegations of academic misconduct each year, placing an undue and unfair burden on students, and causing a climate of anxiety and unfair scrutiny similar to the one produced by proctoring technology during the Covid-19 pandemic [22].

---

[1] While the AI detection score was initially displayed on all assignments by default, after some backlash from UK universities an option to hide it was made available.

*Detectors are incompatible with co-creation and good usage*

After an initial period of stop-gap measures where students were simply told to not use generative AI, there has been a growing recognition of the need to integrate and teach responsible use. This is outlined, for example, in the principles adopted by UK universities in the Russell Group [40], which have acknowledged the significant benefits of this technology in various sectors and are actively working to promote AI literacy among students and staff. The third principle reads: "Universities will adapt teaching and assessment to incorporate the ethical use of generative AI and support equal access", while the fourth reads: "Universities will ensure academic rigour and integrity is upheld".

Any policy that employs AI detection is fundamentally incompatible with teaching ethical, critical usage of large language models. Detectors, by their design, are unable to differentiate good from bad usage. Naturally, not every use of AI should be legitimised, and there exists a spectrum of AI usage in academic settings, with some applications constituting malpractice and others representing legitimate, compliant uses that can enrich the learning experience, which may vary depending on the course and learning objectives. The indiscriminate nature of AI detectors fails to recognize this nuance, treating all AI-assisted work with the same level of suspicion, becoming effectively useless as soon as any use of generative AI is allowed.

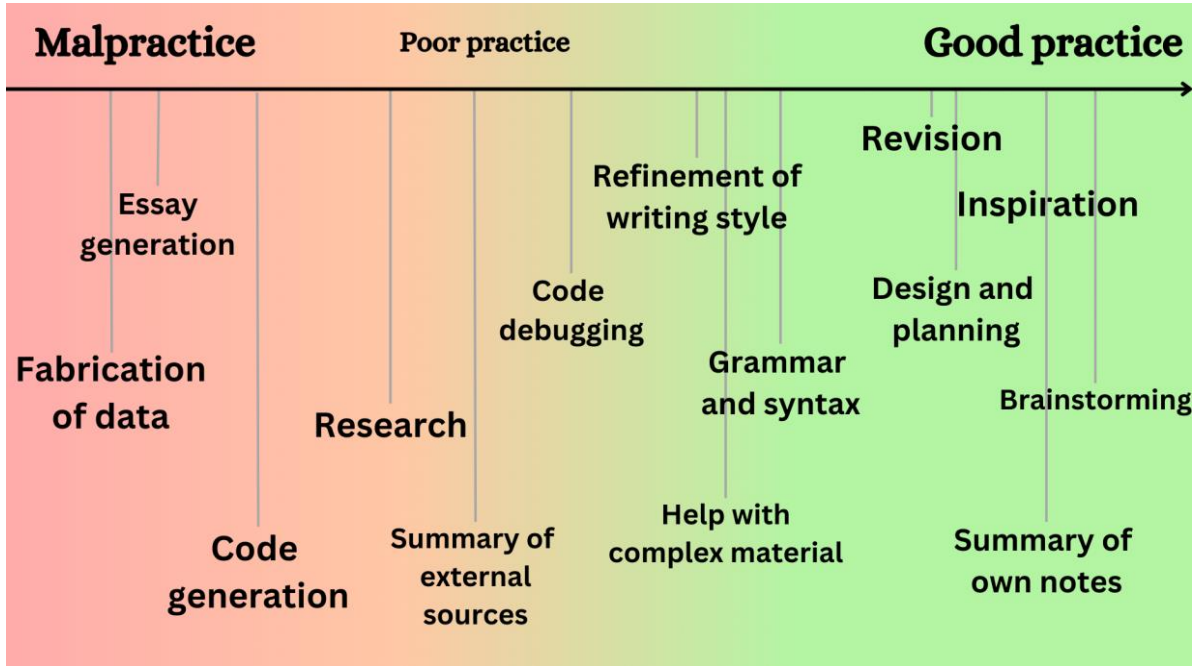

Figure 2: A possible scale of potential uses of generative AI

The result of any detection is to stifle the educational use of a promising new technology. Students, aware that any AI-assisted work might be flagged, might shy away from experimenting with and learning from these tools, even in contexts where their use is encouraged or could offer significant educational benefits. The author has participated in several focus groups on the subject, and the most common concern expressed by students who stated they were not using generative AI was that they would expose themselves to malpractice accusations, and similar reports have recently appeared [32][33][20].

Accessibility technology deserves a separate mention. The visual and audial capabilities of models such as GPT-4 make generative AI a huge asset for inclusivity and accessibility. For instance, generative AI can easily create captions for images, read text out loud from pictures (including handwritten ones), translate text, pictures and videos into different languages, or reword concepts in simpler language [19]. Again the indiscriminate nature of AI detection could constitute a significant barrier for users of accessibility solutions [56].

Should AI detection achieve greater reliability, it might align with policies enforcing a complete ban on generative AI use. However, as soon as institutions encourage, tolerate, or teach the use of generative AI, a fundamental incompatibility with AI detection methods becomes evident, as these tools do not discern between permissible and impermissible uses of AI. The crux of the issue is not merely in the technological limitations of AI detectors but in their philosophical misalignment with the necessities of education. As universities and colleges increasingly and correctly advocate for incorporating generative AI in their curricula, teaching its critical, ethical, balanced and informed usage, the indiscriminate nature of AI detection stands in contrast, pushing in the opposite direction.

*The association between AI and malpractice*

We start by mentioning that correct usage of generative AI has been shown to significantly improve student productivity [13]. In many educational systems, a student's performance is evaluated relative to their peers, with grades adjusted to maintain consistency across different years, classes, and student groups. This implies that the same essay that merited a 70% grade in 2019 might only secure a 60% in 2023 if the widespread use of generative AI among students leads to an overall improvement in the quality of submitted work.

Under such a scaling system, students are therefore implicitly incentivized to utilise generative AI tools. This incentive exists even for those who might not be otherwise inclined to use such technology. The growing demand for AI skills in the job market further complicates this picture. As industries and professions increasingly integrate AI into their operations, the ability to effectively use and understand AI becomes a critical skill.

Educational systems must ensure that all students, regardless of their proficiency with or attitude towards AI, have a fair and equal opportunity to succeed in an environment increasingly influenced by these advanced technologies, while maintaining rigorous standards of academic integrity. As soon as we recognise the vulnerabilities of AI detection, we acknowledge that full-ban policies are inherently unfair.

Even when generative AI usage is permitted by policies, these policies generally tacitly associate generative AI with academic malpractice, which can have unintended consequences. Indeed, such an environment disproportionately affects students who are more cautious or apprehensive about engaging with AI due to the fear of being accused of malpractice. Such a dynamic can exacerbate existing disparities between different groups of students. For example, international students might face more severe consequences if accused of academic malpractice, given the potential implications for their student visas, and hence be more risk-averse.

Conversely, policies need to encourage and promote responsible AI usage from students and incorporate robust, authentic assessment methods that remain effective and equitable regardless of the extent or manner in which AI is utilised, crafted to not only evaluate a student's understanding and skills accurately but also to motivate learning and intellectual engagement.

*Indirect detection and prompt sharing*

In the realm of academic integrity, the strategies for monitoring and assessing students' use of generative AI have evolved, with some institutions considering indirect methods of detection. These methods, while seemingly less intrusive than direct AI detection, are still questionable in their effectiveness and impact on the educational environment, similar to the arguments made above.

This is exemplified in the extremely common pattern in recent generative AI policies in which students are required to share the prompts they used with generative AI tools in assessments[2] [46], [48]. This approach is based on the flawed assumption that generative AI's main usage is as a straightforward input-output mechanism. Instead, in their current mainstream incarnation as conversational chatbots, generative AI tools excel when the user conducts an extensive, iterative dialogue, far exceeding a simple sequence of prompts and responses [52]. Successful, good usage of generative AI may involve hundreds of messages and dozens of threads that may even use different tools. For instance, as acknowledged at the end of this paper, we used GPT-4 [38] for various tasks while writing this paper, and at the time of writing this paragraph the conversation is already over 300 messages long, with complex prompts including entire sections for feedback and revision. The expectation for students to document and disclose the entire interaction is not only taxing and impractical but also conflicts with what good patterns of AI tool use are meant to be. Such a requirement would likely discourage users from fully utilising the potential of this technology.

Another related example is the requirement for students to ring-fence and cite content generated by AI, as they would for an external contribution [12]. While this is certainly appropriate when a paragraph entirely generated by a chatbot is included in any submission, such a policy is entirely impractical for the most common and effective usage of generative AI: co-created content.

A further example is relying on indirect detection methods, such as tracking students' writing patterns through fingerprinting technology (Turnitin [25]), or requiring that students work in a controlled learning environment (Cadmus [6]), reflecting a broader trend of surveillance in education. The idea of tracking the development of students' work through cloud-based platforms, for example, introduces its own set of challenges. While this could ensure that students spend adequate time on their assignments, and provide accurate data to educators, it also imposes a level of surveillance that is often characterised as invasive [30]. Additionally, students could perceive such measures as an implicit expression of mistrust, undermining the student-teacher relationship.

---

[2] A particularly egregious example is the GAIA Policy from the University of Boston [45], that not only requires to add an appendix with the full exchange had with generative AI tools, but goes further in asking students to employ AI detection tools and ensure that their work is not mistakenly flagged (!).

The experience during the COVID-19 pandemic with remote proctoring solutions highlighted the limitations and negative consequences of such surveillance approaches [22]. These methods can lead to heightened anxiety, privacy concerns, and a potential loss of trust between students and educational institutions. Furthermore, their effectiveness is questionable: on one hand, these solutions are also vulnerable to false positives, leading to false accusations of malpractice. On the other hand, students might find ways to circumvent these measures. For instance, AI may soon be able to emulate human usage patterns when using a computer, perhaps using a controlled learning environment bypassing any detection [17].

Ironically, in the current detector-rich environment, students may react positively to the introduction of software that can track drafts and produce an authenticity record of their assignments since that could exonerate them from potential malpractice accusations stemming from misdetection. However, this underscores a concerning trend where the fear of being wrongly accused leads to a willingness to sacrifice certain freedoms – in this case, the freedom to engage in the educational process without constant monitoring. While the intent behind these policies is to uphold academic standards, the resultant environment may foster a culture where surveillance is normalised and autonomy is diminished. This trend is a reminder of the need for careful consideration and balance in implementing technologies and policies, ensuring that they serve to enhance, rather than undermine, the educational experience and the principles of academic freedom and trust.

In particular, the core issue with indirect detection methods is that they attempt to control and monitor the use of AI tools, rather than focusing on adapting educational practices to the new reality of AI. Authentic assessment methods should aim to leverage the potential of AI in enhancing learning while ensuring that essential skills and knowledge are acquired independently of AI assistance. While indirect detection and prompt sharing policies may offer some level of monitoring, they fail to address the fundamental challenge of integrating AI into education in a constructive and non-invasive manner.

## 4. Teach critical usage of AI

*A lesson from the past*

As mathematics lecturers, there are occasions when we wish to assess students on tasks that a calculator, or a tool such as WolframAlpha [53], has been able to perform effortlessly for decades, such as computing the sine or cosine of the sum of two angles using the general formula. When that happens, we create controlled conditions to ensure that calculators cannot be used, normally an in-person in-class test where calculators are not allowed[3]. We would not consider employing post-hoc detection algorithms to try and detect the use of calculators in a submission.

---

[3] There are other strategies: one would be to ask for a reflective task, such as "show your work". Another would be to use numbers that standard calculators are unable to compute, such as very large ones. However, neither of these is particularly useful in our parallel with AI, as a LLM can perform reflective tasks, and is able to handle most inputs that a student would – including visuals or audio files [38].

Generative AI is more disruptive, but not conceptually different. Albeit on a much smaller scale, the integration of calculators into mathematics education provides a valuable historical perspective on the current discussion surrounding the incorporation of generative AI in academic settings.

Initially, the advent of calculators was met with considerable scepticism and resistance within the educational community. Concerns were raised about their potential to diminish traditional learning methods and outcomes [50]. Despite initial bans and apprehensions, students quickly began to adopt calculators, recognizing their immense potential for efficiency and practical application in solving complex mathematical problems. This gradual shift in perception led to a wider acceptance and eventual integration of calculators into the educational curriculum, marking a significant transformation in teaching methodologies [5].

This acceptance, however, was not without thoughtful adaptation. Educational approaches had to be reevaluated and restructured in light of this new technology. Some traditional learning objectives, such as the ability to perform complicated manual calculations of logarithms using tables, essentially disappeared. Others were reshaped to better suit the capabilities of calculators, requiring a deeper understanding of the underlying processes: educators started asking students to explain their methods or solve more intricate problems that stretched beyond the calculators' capacity, such as finding the determinant of an exceptionally large power of a matrix in linear algebra.

Finally, other learning objectives were deemed essential to preserve regardless of calculators, such as fundamental arithmetic skills like integer multiplication taught in primary education. This required adaptation in terms of assessment, where educators designed controlled testing environments where the use of calculators was banned, thereby ensuring the authenticity and integrity of the assessment. This was accomplished without post-hoc detection of calculator usage or the pretence of being able to control what students do outside of the classroom. Instead, the focus was redirected towards creating authentic assessments which ensured integrity and served as a main motivator for students to learn and understand fundamental concepts, regardless of the ease and convenience offered by calculators. Ultimately, calculators are fundamentally integrated into mathematics classes at all levels and are generally regarded as beneficial to students' capabilities and productivity [41].

The introduction and evolving role of generative AI in education calls for a similar, nuanced response. Adapting to this technological advancement involves redefining existing learning outcomes to utilise the unique capabilities of AI. It also calls for introducing more sophisticated and advanced educational objectives that harness the full potential of AI tools.

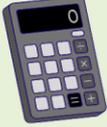

Figure 3: Proposed classification of learning objectives, with some examples

Crucially, in scenarios where independent assessment of skills is necessary, devoid of AI assistance, the establishment of controlled conditions becomes essential, but this approach should not rely on post-hoc detection or the unrealistic expectation of regulating students' use of technology outside the classroom. Instead, it should focus on educational experiences that acknowledge the presence of AI, utilise its benefits, and maintain the essential value and integrity of the educational process. Through such a balanced and forward-looking approach, the educational system can evolve harmoniously with technological progress, embracing new tools and methods while safeguarding the essence of learning and knowledge acquisition.

*Suggestions for effective policies and assessments*

This paper's primary aim is to advocate for the development of robust assessment methods in the era of generative AI, rather than to prescribe specific solutions. The task of fully developing these methods is considerable and complex, likely to be the subject of extensive research and literature in the coming years.

As we have argued, it is imperative to acknowledge and fully embrace the role of generative AI, explicitly integrating it into curricula. This necessitates a clear departure from policies of avoidance or mere tolerance, moving towards proactive and comprehensive incorporation of AI tools in the learning process. Such an approach involves not only recognizing the potential of AI to enhance and transform educational practices but also committing to its integration in a manner that is transparent, well-defined, and aligned with educational objectives. By doing so, institutions can demystify AI for students and faculty alike, paving the way for an informed and constructive engagement, and opening the way for further, more sophisticated approaches. Further, this is crucial to safeguard the role of universities at the peak of the educational landscape, and prepare students for a future where AI is an intrinsic part of professional life.

Incorporating generative AI into educational curricula necessitates a renewed focus on maintaining the human element at the heart of learning. This is crucial not only for preserving human knowledge but also for ensuring that students develop the skills to critically assess AI-generated content. In an era where distinguishing between an advanced AI search engine and a sentient being becomes challenging, the ability to discern and evaluate these outputs becomes a fundamental human skill to safeguard. Flaws and biases in AI outputs are inevitable, as they reflect the limitations of their training data and algorithms: human oversight is essential to validate the output of AI.

The challenge lies in motivating students to learn and engage with material even when AI tools could potentially undertake the task for them. Further to the human element above, authentic assessment methods become key in this scenario. Such an approach aims to cultivate a mindset where students view AI as a supplemental tool rather than a replacement for their knowledge and skills, and AI output is evaluated with a critical perspective [34].

As outlined in the Russell Group principles [40], students should be explicitly taught ethical, principled and critical usage of generative AI, theoretically and practically, contextualised to each of the learning objectives of their courses, and with a shared framework based on mutual trust.

Course leaders should require acknowledgements and transparent disclosures of generative AI usage in each assignment, providing guidelines and examples to do so effectively. Such disclosure fosters an environment of honesty and responsibility, encouraging students to critically reflect on their use of AI and its impact on their learning process, while also enabling educators to appropriately guide and assess students, ensuring that the use of AI enhances rather than diminishes the educational value of their work.

A full U-turn on the progressive reduction of human contact in assessments and courses in general is now essential to safeguard their integrity. For synchronous assessments, such as exams, the value of in-person or invigilated settings cannot be overstated, as ultimately only controlled conditions can enable genuine assessment of certain learning objectives. For asynchronous assessments, like essays or dissertations, incorporating milestones that involve direct human interaction is equally important. Scheduled interactions with a tutor or course leader can serve as significant touchpoints in the authentication process. These interactions, characterised by meaningful conversations and critical feedback, not only help to validate the work as the student's own but also provide opportunities for deeper learning and intellectual development. By automating or significantly speeding up other aspects of the job, such as administrative work, AI can help make more time for authentic, in-person interactions between academics and students.

Alternative methods of assessment beyond traditional written exams and essays can also be effective alternatives to explore: group assessments, for instance, promote collaborative learning and problem-solving skills. By working together, students must negotiate, share ideas, and synthesise their collective knowledge, which encourages them to engage with the material genuinely. Presentations, another alternative, require students to not only understand their subject matter but also communicate it effectively, demonstrating their grasp of the topic in person. Similarly, oral examinations, where students verbally articulate their understanding and reasoning, can offer a dynamic and interactive assessment environment. Peer review exercises, where students evaluate each other's work, foster critical thinking and analytical skills, providing an opportunity for reflective learning. Portfolio assessments, collecting a range of work over a period, can demonstrate a student's progression and depth of understanding across a subject area and can be safeguarded with regular milestones and human connections.

Educators should be wary of underestimating the capabilities of generative AI. Common assertions about AI's limitations in handling scenario-based questions, critical reflection, or complex references should be taken with a grain of scepticism. The landscape of AI is rapidly evolving, with new advancements emerging constantly. For instance, until September 2023 it was common to suggest including visual elements in assessments since available models only allowed textual input. With the release of GPT-4V [38] to the public, a model capable of seeing and processing pictures, that suggestion ceased to be valid. If anything, overestimating the capabilities of generative AI has few drawbacks, and ensures that the developed policies and assessments are more future-proof.

## 5. Conclusion

In light of the analysis and discussion presented, we argue that reliance on detection mechanisms is incompatible with a modern, forward-thinking approach to education in the era of generative AI. The current state of detectors, with insufficient accuracy and susceptibility to basic forms of manipulation, undermines their efficacy. More fundamentally, the nature of detection itself poses inherent challenges. They are not falsifiable, and unable to distinguish between different forms of AI usage, whether compliant or non-compliant with academic standards.

The implications of persisting with detection-oriented policies are far-reaching and predominantly negative. They can contribute to heightened anxiety among students, create an atmosphere of distrust, and potentially delay progress in adapting educational practices to the evolving landscape of AI. These policies would place undue burdens on students, fostering an environment that is counterproductive to both learning and well-being.

Several educational institutions, such as the University of Manchester [47] and Vanderbilt University [9], have already either banned or strongly discouraged the use of AI detection tools, and we encourage all other educational establishments to do the same.

The path forward lies not in clinging to detection methods to preserve antiquated assessment models of the past, but in embracing and adapting to the new paradigm, which is here to stay. This entails fostering an educational environment where AI is recognized as a tool for learning and innovation, and its usage is encouraged. We strongly advocate for the development of robust, authentic assessment methods and policies that encourage responsible AI usage, to ensure that the educational experience remains enriching and equitable, equipping students with the skills they need to become leaders and innovators.

## Acknowledgements

I thank Dr Elliot McKernon for reviewing an early draft of this paper. The following paper has benefitted from OpenAI's GPT-4 model for feedback, rephrasing, grammar suggestions and organisation of content. Do note that the ideas contained in this paper remain the author's original ones. Several arguments made in this paper originally appeared on the author's blog [2].

## References

[1]   D. Adamson, *New research: Turnitin's AI detector shows no statistically significant bias against English Language Learners*, 2023. Available: https://www.turnitin.com/blog/new-

[1] research-turnitin-s-ai-detector-shows-no-statistically-significant-bias-against-english-language-learners. [Accessed: Nov. 19, 2023]

[2] C. G. Ardito, *Against AI Detection*, Thoughts, 2023. Available: https://cesaregardito.substack.com/p/against-ai-detection-1-detection. [Accessed: Nov. 20, 2023]

[3] C. G. Ardito and ChatGPT, *Conversation with ChatGPT 3.5*, (Nov. 2023), Available: https://chat.openai.com/share/666370cd-4333-4b42-88a1-1760e7513c4a. [Accessed: Nov. 19, 2023]

[4] J. L. Borges, *The Library of Babel*, in Ficciones, translated by Anthony Kerrigan, Ed., Grove Press, (1962).

[5] J. Bostic and S. Pape, *Examining Students' Perceptions of Two Graphing Technologies and Their Impact on Problem Solving*, Journal of Computers in Mathematics and Science Teaching, vol. **29**, no. 2, (2010).

[6] Cadmus, *Identifying and Mitigating Risks of AI in Authentic Assessment Practices*, Jan. 2023. Available: https://www.cadmus.io/blog/identifying-and-mitigating-risks-of-ai-in-authentic-assessment-practices. [Accessed: Nov. 22, 2023]

[7] E. Clark, T. August, S. Serrano, N. Haduong, S. Gururangan, and N. A. Smith, *All that's "human" is not gold: Evaluating human evaluation of generated text*, in *ACL-IJCNLP 2021 - 59th Annual Meeting of the Association for Computational Linguistics and the 11th International Joint Conference on Natural Language Processing, Proceedings of the Conference*, 2021. doi: 10.18653/v1/2021.acl-long.565

[8] Cleo Nardo, *Remarks 1–18 on GPT (compressed)*, LessWrong, Mar. 2023. Available: https://www.lesswrong.com/posts/7qSHKYRnqyrumEfbt/remarks-1-18-on-gpt-compressed. [Accessed: Nov. 19, 2023]

[9] M. Coley, *Guidance on AI Detection and Why We're Disabling Turnitin's AI Detector*, Vanderbilt University, Aug. 16, 2023. Available: https://www.vanderbilt.edu/brightspace/2023/08/16/guidance-on-ai-detection-and-why-were-disabling-turnitins-ai-detector/. [Accessed: Nov. 22, 2023]

[10] H. Desaire, A. E. Chua, M. Isom, R. Jarosova, and D. Hua, *Distinguishing academic science writing from humans or ChatGPT with over 99% accuracy using off-the-shelf machine learning tools*, Cell Reports Physical Science, vol. **4**, no. 6, p. 101426, (Jun. 2023), doi: 10.1016/j.xcrp.2023.101426

[11] B. Edwards, *Why AI detectors think the US Constitution was written by AI*, Jul. 2023. Available: https://arstechnica.com/information-technology/2023/07/why-ai-detectors-think-the-us-constitution-was-written-by-ai/. [Accessed: Nov. 19, 2023]

[12] I. Eleftheriou and A. Mubarik, *AI Code of Conduct*, *The University of Manchester*. 2023. Available: https://www.iliada-eleftheriou.com/AICodeOfConduct/. [Accessed: Nov. 22, 2023]

[13]   F. Fauzi, L. Tuhuteru, F. Sampe, A. M. A. Ausat, and H. R. Hatta, *Analysing the Role of ChatGPT in Improving Student Productivity in Higher Education*, Journal on Education, vol. **5**, no. 4, pp. 14886–14891, (Apr. 2023), doi: 10.31004/joe.v5i4.2563

[14]   V. Fishchuk, *Adversarial attacks on neural text detectors*, in *Twente Student Conference on IT*, Jul. 2023.

[15]   V. Fishchuk and D. Braun, *Efficient Black-Box Adversarial Attacks on Neural Text Detectors*, (Nov. 2023).

[16]   G. A. Fowler, *What to do when you're accused of AI cheating*, The Washington Post, Aug. 14, 2023. Available: https://www.washingtonpost.com/technology/2023/08/14/prove-false-positive-ai-detection-turnitin-gptzero/. [Accessed: Nov. 19, 2023]

[17]   N. Friedman, *Using GPT-3 to control a browser (Tweet)*, Sep. 30, 2022. Available: https://twitter.com/natfriedman/status/1575631194032549888. [Accessed: Nov. 22, 2023]

[18]   C. A. Gao et al., *Comparing scientific abstracts generated by ChatGPT to original abstracts using an artificial intelligence output detector, plagiarism detector, and blinded human reviewers*, bioRxiv, vol. **12**, (2022).

[19]   K. S. Glazko et al., *An Autoethnographic Case Study of Generative Artificial Intelligence's Utility for Accessibility*, in *The 25th International ACM SIGACCESS Conference on Computers and Accessibility*, New York, NY, USA: ACM, Oct. 2023, pp. 1–8. doi: 10.1145/3597638.3614548

[20]   T. Gorichanaz, *Accused: How students respond to allegations of using ChatGPT on assessments*, Learning: Research and Practice, vol. **9**, no. 2, pp. 183–196, (Jul. 2023), doi: 10.1080/23735082.2023.2254787

[21]   J. Gullifer and G. A. Tyson, *Exploring university students' perceptions of plagiarism: a focus group study*, Studies in Higher Education, vol. **35**, no. 4, pp. 463–481, (Jun. 2010), doi: 10.1080/03075070903096508

[22]   D. Harwell, *Cheating-Detection Companies Made Millions During the Pandemic. Now Students Are Fighting back \**, in Ethics of Data and Analytics, Boca Raton: Auerbach Publications, (2022), pp. 410–417. doi: 10.1201/9781003278290-60

[23]   M. Heikkilä, *How AI-generated text is poisoning the internet*, MIT Technology Review, Dec. 16, 2022.

[24]   J. M. C. Hughes and D. L. McCabe, *Understanding Academic Misconduct*, Canadian Journal of Higher Education, vol. **36**, no. 1, (2006), doi: 10.47678/cjhe.v36i1.183525

[25]   T. S. Hui and D. Ng, *How can Singapore's universities deter AI-assisted cheating in the age of ChatGPT?*, *Channel News Asia*, Feb. 27, 2023. Available: https://www.channelnewsasia.com/singapore/chatgpt-openai-chatbot-education-universities-schools-students-3309201. [Accessed: Nov. 22, 2023]

[26] M. Jakesch, J. T. Hancock, and M. Naaman, *Human heuristics for AI-generated language are flawed*, Proceedings of the National Academy of Sciences of the United States of America, vol. **120**, no. 11, (2023), doi: 10.1073/pnas.2208839120

[27] M. Klee, *She Was Falsely Accused of Cheating With AI — And She Won't Be the Last*, Rolling Stone, Jun. 06, 2023. Available: https://www.rollingstone.com/culture/culture-features/student-accused-ai-cheating-turnitin-1234747351/. [Accessed: Nov. 19, 2023]

[28] N. Köbis and L. D. Mossink, *Artificial intelligence versus Maya Angelou: Experimental evidence that people cannot differentiate AI-generated from human-written poetry*, Computers in Human Behavior, vol. **114**, (2021), doi: 10.1016/j.chb.2020.106553

[29] K. Krishna, Y. Song, M. Karpinska, J. Wieting, and M. Iyyer, *Paraphrasing evades detectors of AI-generated text, but retrieval is an effective defense*, (Mar. 2023).

[30] V. Lenard, *USyd Business School quietly trials assessment platform Cadmus*, Honi Soit, Aug. 14, 2023. Available: https://honisoit.com/2023/08/usyd-business-school-quietly-trials-assessment-platform-cadmus/. [Accessed: Nov. 22, 2023]

[31] W. Liang, M. Yuksekgonul, Y. Mao, E. Wu, and J. Zou, *GPT detectors are biased against non-native English writers*, *Patterns*, vol. **4**, no. 7. 2023. doi: 10.1016/j.patter.2023.100779

[32] U. Litvinaite, *Academic integrity and assessment in the context of digitalisation and the rise of generative AI: student perspective*, (2023). Available: https://educational-innovation.sydney.edu.au/teaching@sydney/students-answer-your-questions-about-generative-ai-part-2-ethics-integrity-and-the-value-of-university/. [Accessed: Nov. 20, 2023]

[33] D. Liu and A. Bridgeman, *Students answer your questions about generative AI – part 2: Ethics, integrity, and the value of university*, *Teaching@Sydney*, 2023.

[34] J. M. Lodge, S. Howard, M. Bearman, and P. Dawson, *Assessment reform for the age of Artificial Intelligence.*, (2023).

[35] A. R. Malik *et al.*, *Exploring Artificial Intelligence in Academic Essay: Higher Education Student's Perspective*, International Journal of Educational Research Open, vol. **5**, p. 100296, (Dec. 2023), doi: 10.1016/j.ijedro.2023.100296

[36] J. Monteiro, F. Silva-Pereira, and M. Severo, *Investigating the existence of social networks in cheating behaviors in medical students*, BMC Medical Education, vol. **18**, no. 1, (2018), doi: 10.1186/s12909-018-1299-7

[37] OpenAI, *New AI classifier for indicating AI-written text*, Jul. 2023. Available: https://openai.com/blog/new-ai-classifier-for-indicating-ai-written-text. [Accessed: Nov. 19, 2023]

[38] OpenAI, *GPT-4 Technical Report*, (Mar. 2023).

[39] R. F. Parks, P. B. Lowry, R. T. Wigand, N. Agarwal, and T. L. Williams, *Why students engage in cyber-cheating through a collective movement: A case of deviance and collusion*, Computers and Education, vol. **125**, (2018), doi: 10.1016/j.compedu.2018.04.003

[40] Russell Group, *New principles on use of AI in education*, (Jul. 2023), Available: https://russellgroup.ac.uk/news/new-principles-on-use-of-ai-in-education/. [Accessed: Nov. 20, 2023]

[41] L. M. Simonsen and T. P. Dick, *Teachers' perceptions of the impact of graphing calculators in the mathematics classroom*, Journal of Computers in Mathematics and Science Teaching, vol. **16**, no. 2, (1997).

[42] D. Sokol, *It is too easy to falsely accuse a student of using AI: a cautionary tale*, Times Higher Education, Jul. 10, 2023.

[43] Turnitin, *AI Writing Detection Capabilities F.A.Q.*, 2023. Available: https://www.turnitin.com/products/features/ai-writing-detection. [Accessed: Nov. 19, 2023]

[44] Turnitin, *Interpreting the Similarity Report*, 2023. Available: https://help.turnitin.com/feedback-studio/turnitin-website/instructor/the-similarity-report/interpreting-the-similarity-report.htm. [Accessed: Nov. 20, 2023]

[45] University of Boston, *CDS Generative AI Assistance (GAIA) Policy*, 2023. Available: https://www.bu.edu/cds-faculty/culture-community/gaia-policy/. [Accessed: Nov. 22, 2023]

[46] University of Huddersfield, *AI Generated Text*, 2023. Available: https://library.hud.ac.uk/pages/referencing-aitext/. [Accessed: Nov. 22, 2023]

[47] University of Manchester, *Artificial Intelligence (AI) Teaching Guidance*, Oct. 2023.

[48] University of Melbourne, *Acknowledging, citing and referencing use of AI tools and technologies*, 2023. Available: https://students.unimelb.edu.au/academic-skills/resources/referencing/acknowledging-use-of-ai-tools-and-technologies. [Accessed: Nov. 22, 2023]

[49] J. Vincent, *Meta's powerful AI language model has leaked online — what happens now?*, (Mar. 2023), Available: https://www.theverge.com/2023/3/8/23629362/meta-ai-language-model-llama-leak-online-misuse. [Accessed: Nov. 19, 2023]

[50] B. K. Waits and F. Demana, *Calculators in mathematics teaching and learning: Past, present, and future.*, in Learning mathematics for a new century, (2000).

[51] D. Weber-Wulff *et al.*, *Testing of detection tools for AI-generated text*, International Journal for Educational Integrity, vol. **19**, no. 1, p. 26, (Dec. 2023), doi: 10.1007/s40979-023-00146-z

[52] S. Willison, *Think of language models like ChatGPT as a "calculator for words,"* Weblog, Apr. 02, 2023. Available: https://simonwillison.net/2023/Apr/2/calculator-for-words/. [Accessed: Nov. 23, 2023]

[53] Wolfram Research, *Wolfram Alpha*. 2023. Available: https://www.wolframalpha.com/. [Accessed: Nov. 20, 2023]

[54] J. Wu, S. Yang, R. Zhan, Y. Yuan, D. F. Wong, and L. S. Chao, *A Survey on LLM-generated Text Detection: Necessity, Methods, and Future Directions*, (Oct. 2023).

[55] H. Zhang, B. L. Edelman, D. Francati, D. Venturi, G. Ateniese, and B. Barak, *Watermarks in the Sand: Impossibility of Strong Watermarking for Generative Models*, (Nov. 2023).

[56] *AI & Accessibility*, Cornell University, Center For Teaching Innovation, 2023. Available: https://teaching.cornell.edu/generative-artificial-intelligence/ai-accessibility. [Accessed: Nov. 20, 2023]